\documentclass[pra,aps,superscriptaddress,twocolumn,color]{revtex4-1}

\usepackage{amsmath}
\usepackage{amsfonts}
\usepackage{bm}
\usepackage{graphicx}
\usepackage{array,color}

\begin{document}
\title{Vortex pairs in a spin-orbit coupled Bose-Einstein condensate}

\author{Masaya Kato}
\affiliation{Department of Engineering Science, University of Electro-Communications, Tokyo 182-8585, Japan}

\author{Xiao-Fei Zhang}
\affiliation {Key Laboratory of Time and Frequency Primary Standards, National Time Service Center, Chinese Academy of Sciences, Xi'an 710600, China}
\affiliation {University of Chinese Academy of Sciences, Beijing 100049, China}

\author{Hiroki Saito}
\affiliation {Department of Engineering Science, University of Electro-Communications, Tokyo 182-8585, Japan}

\date{\today}
\begin{abstract}

Static and dynamic properties of vortices in a two-component Bose-Einstein
condensate with Rashba spin-orbit coupling are investigated.
The mass current around a vortex core in the plane-wave phase is found to
be deformed by the spin-orbit coupling, and this makes the dynamics of
the vortex pairs quite different from those in a scalar Bose-Einstein
condensate.
The velocity of a vortex-antivortex pair is much smaller than that without
spin-orbit coupling, and there exist stationary states.
Two vortices with the same circulation move away from each other or unite
to form a stationary state.

\end{abstract}


\maketitle

\section{Introduction}
\label{s:introduction}

Topological excitations in superfluids originate from the intertwining
between internal and external degrees of freedom in the order
parameters.
The simplest example is a quantized vortex in a scalar superfluid, in
which the complex order parameter with the U(1) manifold winds around the
vortex core, producing azimuthal superflow~\cite{Onsager,Feynman}.
For the order parameters with spin degrees of freedom, a rich variety of
topological excitations are possible; these include
skyrmions~\cite{Leslie,Choi}, monopoles~\cite{Ray}, 
half-quantum vortices~\cite{Seo}, and knots~\cite{Hall}.
Because of the close relationship between the spin and motional degrees of
freedom in the topological excitations, we expect that their static and
dynamic properties are significantly altered if there exists coupling
between them, that is, if there exists spin-orbit coupling (SOC).

Recently, Bose-Einstein condensates (BECs) of ultracold atomic gases with
SOC have been realized experimentally~\cite{Lin,Zhang,Campbell,Wu,JLi};
in these experiments, the atomic spin or quasispin was coupled with the
atomic momentum using Raman laser beams.
Numerous theoretical studies have been performed to evaluate the static
properties of topological excitations in spin-orbit (SO) coupled BECs,
e.g., vortex arrays~\cite{Wang},
vortices in rotating systems~\cite{XQXu,Radic,Zhou,Liu},
half-quantum vortices~\cite{Sinha,Lobanov},
skyrmions~\cite{Hu,Kawakami,CFLiu,Chen,YLi},
topological spin textures~\cite{Kawakami2,Xu,Ruo,Han,Han2},
dipole-induced topological structures~\cite{Deng,Wilson,Kato},
and solitons with vortices~\cite{Sakaguchi,YCZhang}.
However, there have been only a few studies on their dynamics in
SO-coupled BECs.
The dynamics of a single quantized vortex in a harmonic trap was
considered in Refs.~\cite{Fetter,Kasamatsu}.

In this paper, we investigate the dynamics of a quantized vortex pair in a
quasispin-$1/2$ BEC with Rashba SOC.
When a singly quantized vortex is created in a uniform plane-wave state,
the phase distribution around the vortex core is significantly altered by
the SOC; this indicates that the mass current around the vortex is quite
different from that without SOC and affects the dynamics of a vortex
pair.
As a result, a vortex-antivortex pair will be stationary or will travel
much more slowly than one without SOC.
The dynamics of a vortex-vortex pair with the same circulation are also
quite different from those without SOC; the vortices move away from each
other, or they approach each other and unite to form a stationary state.

This paper is organized as follows.
The problem is formulated in Sec.~\ref{s:formulation}.
The static properties of a single vortex are studied in Sec.~\ref{s:SV}.
The dynamics of a vortex-antivortex pair and those of a vortex-vortex pair
with the same circulation are investigated in Secs.~\ref{s:VAVP} and
~\ref{s:VVP}, respectively.
Conclusions are presented in Sec.~\ref{s:conclusions}.

\section{Formulation of the problem}
\label{s:formulation}
We consider a two-dimensional (2D) quasispin-1/2 BEC in a uniform system
with Rashba SOC.
Within the framework of mean-field theory, the system can be described by
the order parameter $\bm\Psi(\bm r)=[\psi_{1}(\bm r),\psi_{2}(\bm r)]^T$,
where $T$ denotes the transpose.
The kinetic and SOC energies are given by
\begin{equation}
\label{eq:Ekin}
E_0 \left[\bm \Psi \right]=\int d \bm r \bm\Psi^\dagger \left(
\frac{\bm{p}^2}{2m} - \frac{\hbar k_0}{m} \bm{p} \cdot \bm{\sigma}_\perp
\right)  \bm\Psi,
\end{equation}
where $m$ is the atomic mass, $k_0$ is the strength of the SOC,
and $\bm{\sigma}_\perp = (\sigma_x, \sigma_y)$ are the $2\times 2$ Pauli
matrices. The $s$-wave contact interaction energy is written as
\begin{equation}
 \label{eq:Eint}
E_{\rm int}\left[\bm \Psi \right] = \int d \bm r\left(\frac{\textmd{g}_0}{2}
\sum_{j=1}^{2} |\psi_j|^4 + \textmd{g}_{12} |\psi_1|^2|\psi_2|^2 \right),
\end{equation}
where $\textmd{g}_0$ and $\textmd{g}_{12}$ are the intra- and
inter-component interaction coefficients, respectively.
The total energy is given by
\begin{equation}
\label{eq:E}
E[\bm \Psi] = E_0[\bm \Psi] + E_{\rm int}[\bm \Psi].
\end{equation}

In this paper, we consider an infinite system in which the atomic density
$\bm{\Psi}^\dagger \bm{\Psi}$ far from vortices is a constant, $n_0$.
In the following, we normalize the length, velocity, time, and energy by
the healing length $\hbar/ \sqrt{m \textmd{g}_0 n_0}$, the sound
velocity $\sqrt{\textmd{g}_0 n_0 /m}$, the characteristic time scale
$\hbar/(\textmd{g}_0 n_0)$, and the chemical potential
$\textmd{g}_0 n_0$.
The dimensionless coupled Gross-Pitaevskii (GP) equations, $i \partial
\bm{\Psi} / \partial t = \delta E[\bm{\Psi}] / \delta \bm{\Psi}$, have the
form
\begin{subequations} \label{GPE}
\begin{equation}
\label{eq:GPE1}
i\frac{\partial\psi_1}{\partial t} = -\frac{1}{2}\bm{\nabla}^2\psi_1+i\kappa\partial_{-}\psi_2
+\left(|\psi_1|^2+\gamma|\psi_2|^2 \right)\psi_1,
\end{equation}
\begin{equation}
\label{eq:GPE2}
i\frac{\partial\psi_2}{\partial t} = -\frac{1}{2}\bm{\nabla}^2\psi_2+i\kappa\partial_{+}\psi_1
+\left(\gamma|\psi_1|^2+|\psi_2|^2 \right)\psi_2,
\end{equation}
\end{subequations}
where $\partial_{\pm}=\partial/\partial x \pm i \partial/\partial y$,
$\kappa = \hbar k_0/\sqrt{m \textmd{g}_0 n_0}$, and the ratio between the
inter- and intra-component interactions is $\gamma =
\textmd{g}_{12}/\textmd{g}_0$.
The ground state is the plane-wave state for $\gamma < 1$ and the stripe
state for $\gamma > 1$~\cite{Wang}, which breaks the rotational symmetry
of the system.
In the following discussion, we will focus on the miscible case, $\gamma <
1$, and the ground state is given by the plane-wave state,
\begin{equation}
\label{eq:PL}
\bm \Psi(\bm r) =
\frac{1}{\sqrt{2}}
\left(
\begin{array}{c}
e^{i \kappa x}\\
e^{i \kappa x}
\end{array}
\right),
\end{equation}
where the wave vector is chosen to be in the $x$ direction.

The velocity field is useful for understanding the dynamics of vortices.
From the equation of continuity $\partial \rho / \partial t + \bm \nabla
\cdot (\rho \bm v) = 0$ with atomic density $\rho = |\psi_1|^2 +
|\psi_2|^2$, we obtain the velocity field as
\begin{eqnarray}
v_\xi(\bm r) & = & \frac{1}{2i \rho(\bm{r})}\left[
\bm{\Psi}^\dagger(\bm{r}) \nabla_\xi \bm{\Psi}(\bm{r}) -
\bm{\Psi}(\bm{r})^T \nabla_\xi \bm{\Psi}^*(\bm{r}) \right]
\nonumber \\
& & - \kappa S_\xi(\bm{r}), \;\;\;\;\; (\xi = x, y)
\label{eq:VF}
\end{eqnarray}
where
\begin{equation} \label{S}
S_\xi(\bm{r}) = \frac{1}{\rho(\bm{r})}
\bm{\Psi(\bm{r})}^\dagger \sigma_\xi \bm{\Psi(\bm{r})}
\;\;\;\;\; (\xi = x, y, z)
\end{equation}
is the pseudospin density.
The first term in Eq.~(\ref{eq:VF}) corresponds to the canonical part
related to the superfluid velocity, and the second term corresponds to the
gauge part induced by the SOC.
The velocity field vanishes for the vortex-free ground state in
Eq.~(\ref{eq:PL}), since the first and second terms in Eq.~(\ref{eq:VF})
cancel each other.

We numerically solve Eq.~(\ref{GPE}) by the pseudospectral method with
the fourth-order Runge-Kutta scheme.
In the imaginary-time propagation, on the left-hand side of
Eq.~(\ref{GPE}), $i$ is replaced with $-1$.
The numerical space is taken to be $400 \times 400$, which is sufficiently
large, and the effect of the periodic boundary condition can be neglected.

\section{Single vortex}
\label{s:SV}

\begin{figure}[tb]
\includegraphics[width=8.5cm]{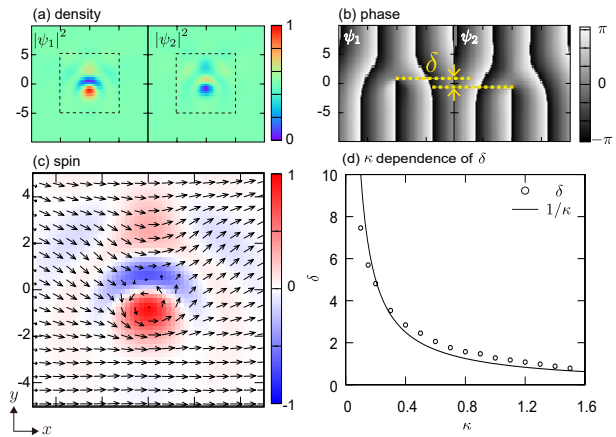}
\caption{(a)-(c) Stable stationary state of a single vortex with
counterclockwise circulation for $\kappa=1$ and $\gamma = 0.8$.
Panels (a) and (b) show the density and phase profiles of each component,
where the unit of density is $n_0$.
In (b), $\delta$ is the distance between the phase defects in the two
components.
Panel (c) shows the spin distribution $\bm{S}(\bm{r})$ defined in
Eq.~(\ref{S}).
The arrows indicate the transverse direction of the spin vector, and the
background color indicates the value of $S_z$.
The dashed square region in (a) is shown magnified in (c).
(d) $\kappa$ dependence of the vortex shift $\delta$.
The solid curve shows $1 / \kappa$ for comparison.}
\label{f:SV}
\end{figure}

We begin with a single vortex state, in which each component contains a
singly quantized vortex.
The initial state of the imaginary-time propagation is
\begin{equation}
\label{eq:SV_INIT}
\bm \Psi(\bm r) =
\frac{1}{\sqrt{2}}
\left(
\begin{array}{c}
e^{i[\Phi(\bm r) + \kappa x]}\\
e^{i[\Phi(\bm r) + \kappa x]}
\end{array}
\right),
\end{equation}
where $\Phi(\bm r) = \tan^{-1}(y/x)$.
After sufficiently long imaginary-time propagation, we obtain the stable
stationary state, as shown in Fig.~\ref{f:SV}.
Figures~\ref{f:SV}(a) and \ref{f:SV}(b) show the density and phase
distributions of the stationary state.
We note that the phase defect in component 1 (2) is shifted in the
$+y$ ($-y$) direction.
We define the distance between the phase defects as $\delta$.
The vortex core in each component is occupied by the other component.
This structure can therefore be regarded as a pair of half-quantum
vortices; nevertheless, we will refer to it as a ``single vortex'' in this
paper.
In the absence of SOC, such a pair of half-quantum vortices repel each
other and cannot form a stationary state~\cite{Eto}.
A similar structure is also found in a one-dimensional SOC
system~\cite{Kasamatsu}.
Figure~\ref{f:SV}(c) shows the spin distribution, and we can see a spin
vortex near the origin.
The dependence of the vortex shift $\delta$ on the SOC strength $\kappa$
is shown in Fig.~\ref{f:SV}(d), which implies $\delta \simeq 1 / \kappa$.

\begin{figure}[tb]
\includegraphics[width=8.5cm]{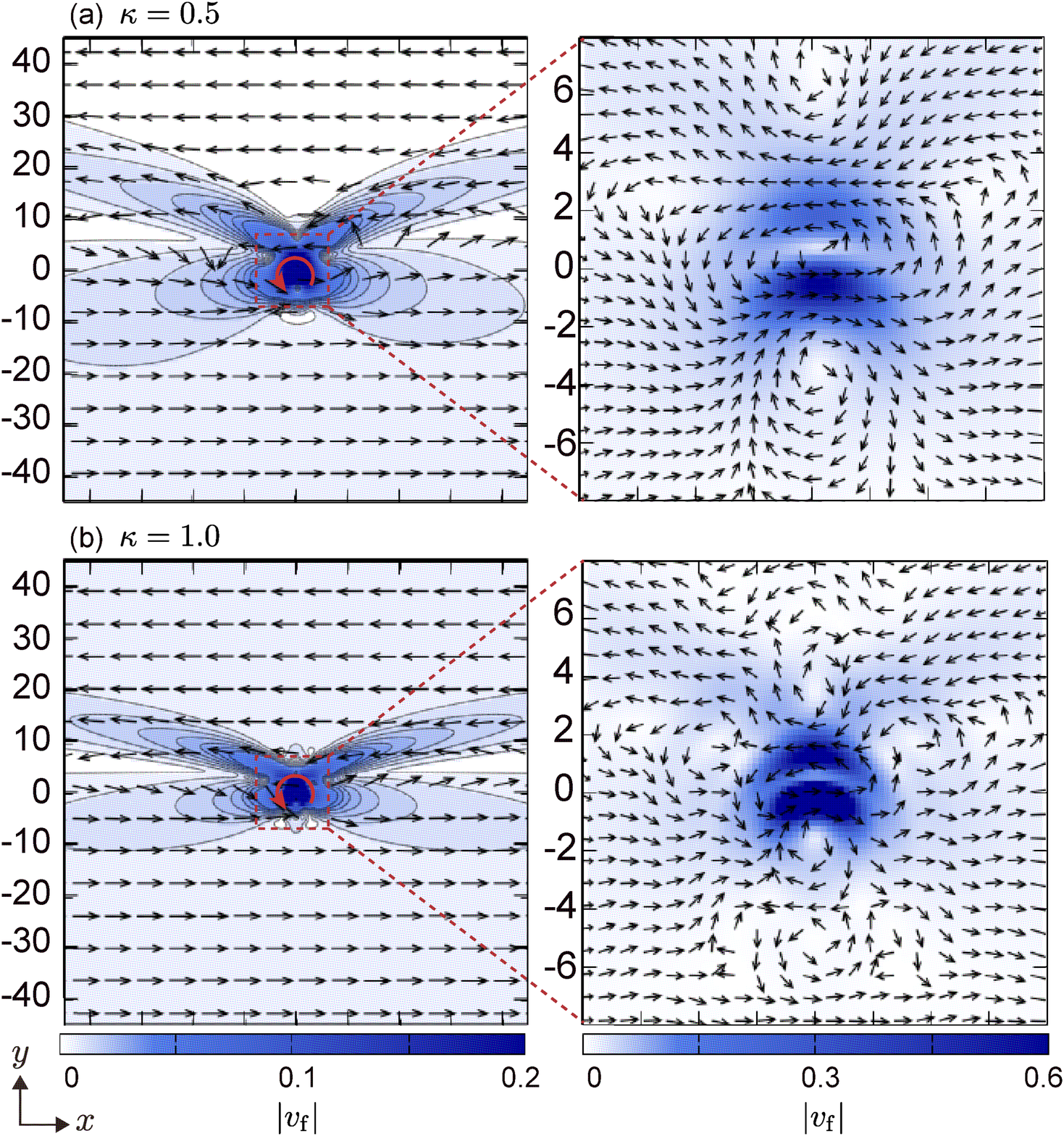}
\caption{Velocity field $\bm{v}(\bm{r})$ of the single-vortex state for
$\gamma = 0.8$ and (a) $\kappa=0.5$ and (b) $\kappa=1$.
The arrows indicate the directions of the velocity, and the background
color indicates the value of $|\bm v|$.
The regions in the dashed squares are magnified in the right-hand panels.
The red arrows indicate the direction of the vortex.
}
\label{f:VF}
\end{figure}

Figure~\ref{f:VF} shows the velocity field $\bm{v}(\bm{r})$ of the
single-vortex state.
The velocity field is greatly deformed by the SOC, compared with the
rotationally symmetric velocity field without SOC.
We note that the deformation of the velocity field extends over a wide
range, and the upper region ($y \gtrsim 10$) exhibits a uniformly leftward
velocity field, while the lower region ($y \lesssim 10$) is rightward.
In these regions, $|\bm{v}| \lesssim 0.01$, which is much smaller than
that without SOC, $|\bm{v}| = 1 / r$.
This effect of SOC is also seen in Fig.~\ref{f:SV}(b), where the phase in
the upper and lower regions is almost $\propto e^{i \kappa x}$, i.e., the
$2\pi$ phase rotation around the vortex core is strongly compressed
around the $x$-axis.
The velocity field near the vortex core exhibits complicated structures
containing multiple circulations, as shown in the right-hand panels in
Fig.~\ref{f:VF}.

Due to the symmetry of the GP equation in Eq.~(\ref{GPE}), the
single-vortex state with clockwise circulation can be obtained from that
with counterclockwise circulation by the following transformation:
\begin{equation} \label{invert}
\psi_1(x, y) \rightarrow \psi_2(x, -y), \;\;\;
\psi_2(x, y) \rightarrow \psi_1(x, -y).
\end{equation}
By this transformation, the winding number of the vortex is inverted
without changing the direction of the plane wave $e^{i\kappa x}$.
Applying the transformation to the state shown in Fig.~\ref{f:SV}, we find
that the vortex core in component 1 (2) shifts in the $+y$ ($-y$) direction
also for the clockwise vortex.
The velocity field and the pseudospin density are transformed as $v_x(x,
y) \rightarrow v_x(x, -y)$, $v_y(x, y) \rightarrow -v_y(x, -y)$, $S_x(x,
y) \rightarrow S_x(x, -y)$, and $S_y(x, y) \rightarrow -S_y(x, -y)$.

For a better understanding of the numerical results, we perform
variational analysis. 
The variational wave function is
\begin{equation}
\label{eq:var}
\bm \Psi(\bm r) =
\frac{1}{\sqrt{2}}
\left(
\begin{array}{c}
e^{i[\Phi_1(\bm r) + \kappa x]}\\
e^{i[\Phi_2(\bm r) + \kappa x]}
\end{array}
\right).
\end{equation}
Substitution of this wave function into Eq.~(\ref{eq:Ekin}) yields
\begin{eqnarray}
\label{eq:model1_E}
E_0 & = & \int d\bm{r} \biggl[ \frac{1}{2} (\bm{\nabla} \chi)^2 +
\frac{1}{8} (\bm{\nabla} \phi)^2
+ \kappa \frac{\partial \chi}{\partial x}
\nonumber \\
& & -\kappa \left( \frac{\partial \chi}{\partial x} + \kappa \right)
\cos\phi + \kappa \frac{\partial \chi}{\partial y} \sin\phi \biggr],
\end{eqnarray}
where $\chi = (\Phi_1 + \Phi_2) / 2$, $\phi = \Phi_1 - \Phi_2$, and the
constant term is neglected.
The first and second lines in Eq.~(\ref{eq:model1_E}) correspond to the
kinetic and SOC energies, respectively.
From the numerical results that the cores are shifted by $\delta$ and that
the $2\pi$ phase rotation around the vortex core is compressed in the $y$
direction, the phases in Eq.~(\ref{eq:var}) are assumed to be
\begin{subequations}
\begin{eqnarray}
\label{eq:model2}
\Phi_1(\bm r) = \tan^{-1}\left( \lambda \frac{y-\delta_y/2}{ x -\delta_x/2} \right),\\
\Phi_2(\bm r) = \tan^{-1}\left( \lambda \frac{y+\delta_y/2}{ x +\delta_x/2} \right),
\end{eqnarray}
\end{subequations}
where $\lambda$ and $\bm{\delta}$ are variational parameters.
We substitute these phases into Eq.~(\ref{eq:model1_E}) and integrate with
respect to $\theta$.
Because of the complicated structure near the vortex cores, we consider
the region in which $r \gg 1$.
The energy is
\begin{eqnarray}
E_0 & = & \int r dr \Biggl\{ -2\pi\kappa^2 + \frac{\pi}{2\lambda r^2}
\nonumber \\
& & + \frac{\pi \kappa^2}{2r^2} \left[ \delta_x^2+ \lambda^2
\left(\delta_y-\frac{1}{\kappa}\right)^2\right] + O(r^{-3}) \Biggr\},
\label{e0}
\end{eqnarray}
which is minimized by $\delta_x = 0$ and $\delta_y = 1/\kappa$.
Thus, the energy is lowered by the displacement of the vortex cores in the
$y$ direction, and the displacement $\delta_y$ is estimated to be $1 /
\kappa$; this is in good agreement with the numerical results shown in
Fig.~\ref{f:SV}(d).
The energy in Eq.~(\ref{e0}) decreases as $\lambda$ increases, and this
accounts for the compressed $2\pi$ phase rotation.
A better variational wave function will allow us to determine the value of
$\lambda$.
We note that the term $\propto \delta_y$ in Eq.~(\ref{e0}) originates from
the last term in the integrand of Eq.~(\ref{eq:model1_E}), which thus
plays an important role in the vortex deformation due to the SOC.

\section{Vortex pair}

\begin{figure}[tb]
\includegraphics[width=8.0cm]{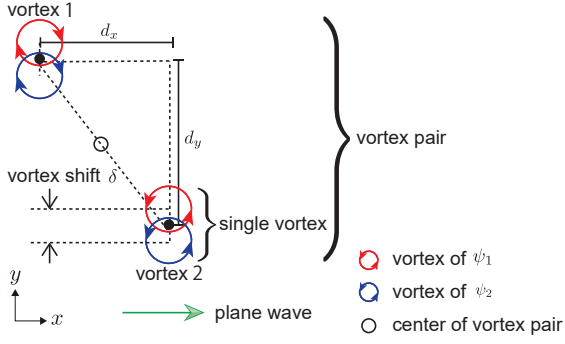}
\caption{
Schematic illustration of the vortex pair for $\langle 1,-1 \rangle$.
The black points are the vortex positions $(x_1,y_1)$ and $(x_2,y_2)$,
which are defined as the midpoints between the phase defects in the two
components.
The open circle is the center of the vortex pair $(x_c,y_c) = ((x_1 +
x_2) / 2, (y_1 + y_2) /2)$.
}
\label{f:schema}
\end{figure}

First, for clarity, we define the positions of the vortices and the
distances between them, as shown in Fig.~\ref{f:schema}.
The position of the phase defect of the $j$th vortex in component $i$ is
denoted by $(x_{ij}, y_{ij})$.
As shown in Fig.~\ref{f:SV}(b), in each vortex, the cores in the two
components are shifted by $\delta$ in the $y$ direction, and then $x_{1j}
= x_{2j}$ and $y_{1j} - y_{2j} = \delta$.
The position of the single vortex is defined by $(x_j, y_j) = ((x_{1j} +
x_{2j}) / 2, (y_{1j} + y_{2j}) / 2)$.
For a vortex pair, the index $j$ is taken in such a way that $y_1 > y_2$.
The distance between the vortices is defined by
$(d_x, d_y) = (x_1 - x_2, y_1 - y_2)$ and $d = (d_x^2 + d_y^2)^{1/2}$.
The center of the vortex pair is defined by $(x_c,y_c) = ((x_1 + x_2) / 2,
(y_1 + y_2) /2)$.
The winding numbers of the first and second vortices are denoted by
$\langle n_1, n_2 \rangle$.
In the following subsections, we will consider the vortex pairs $\langle
\pm 1, \mp 1\rangle$ and $\langle \pm 1, \pm 1\rangle$, which we call
vortex-antivortex pairs and vortex-vortex pairs, respectively.

\subsection{Vortex-antivortex pair}
\label{s:VAVP}

In the absence of SOC, a vortex-antivortex pair travels at a constant
velocity or is annihilated~\cite{Aioi,C. Rorai}.
A vortex-antivortex pair is stationary only in a trap
potential~\cite{Crasovan}, and there is no stationary state in a uniform
system.

\begin{figure}[tb]
\includegraphics[width=8.5cm]{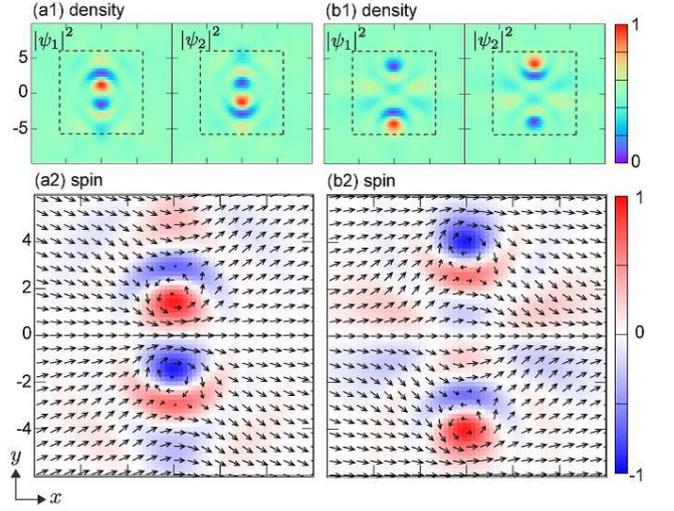}
\caption{Stable stationary states of vortex-antivortex pairs for
$\kappa = 1$ and $\gamma = 0.8$.
The winding combinations are (a) $\langle 1, -1\rangle$ and (b)
$\langle -1, 1\rangle$.
Panels (a1) and (b1) show the density profiles, and (a2) and (b2) show the
spin distributions.
The regions indicated by dashed squares in (a1) and (b1) correspond to
(a2) and (b2), respectively.
The arrows indicate the directions of the transverse spin vector, and
the background color indicates the value of $S_z$.}
\label{f:VAVP}
\end{figure}

In the presence of the SOC, our numerical results show that stable
stationary vortex-antivortex pairs can be formed with a proper choice of
the distance between vortices $d$; an example is shown in
Fig.~\ref{f:VAVP}.
We prepare the initial state in Eq.~(\ref{eq:SV_INIT}) with
\begin{equation} \label{Phi}
\Phi(\bm{r}) = \sum_{j = 1}^2
n_j \tan^{-1} \frac{y - y_j}{x - x_j},
\end{equation}
where, for this example, $n_1 = \pm 1$, $n_2 = \mp 1$, $x_1 = x_2 = 0$, and
$y_1 = -y_2 = d_i / 2$ with $d_i$ being the initial distance between
vortices.
From this initial state, the imaginary-time propagation is performed
sufficiently.
The stationary state is always reached if the initial distance is $d_i
\gtrsim 20$.
It can be seen in Fig.~\ref{f:VAVP} that the distance between vortices in
the stationary state is $d_y \simeq 4.1$ for $\langle 1, -1 \rangle$ and 
$d_y \simeq 6.5$ for $\langle -1, 1 \rangle$.

\begin{figure}[tb]
\includegraphics[width=8.5cm]{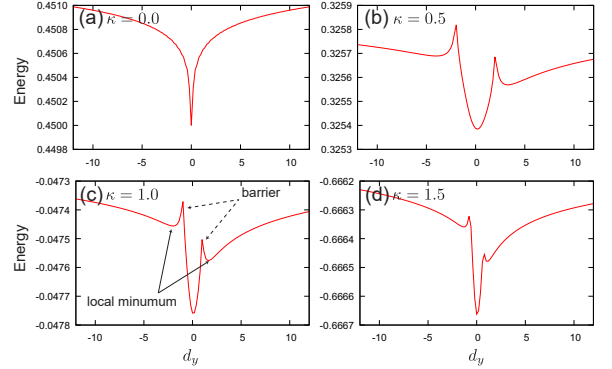}
\caption{
Total energy of a vortex-antivortex pair as a function of the distance
between vortices $d_y$ for (a) $\kappa=0$, (b) $\kappa=0.5$, (c)
$\kappa=1$, and (d) $\kappa=1.5$.
In (b)-(d), local energy minima appear; these correspond to the stationary
states shown in Fig.~\ref{f:VAVP}.
}
\label{f:M3E}
\end{figure}

To understand the stabilization mechanism of the stationary
vortex-antivortex pairs, we calculate the total energy using a model
function given by
\begin{equation}
\label{eq:model3}
\bm \Psi(\bm r) =
\left(
\begin{array}{c}
\sqrt{\rho_1(\bm r)} e^{i[\Phi_1(\bm r) + \kappa x]}\\
\sqrt{\rho_2(\bm r)} e^{i[\Phi_2(\bm r) + \kappa x]}
\end{array}
\right),
\end{equation}
with phases
\begin{subequations}
\label{eq:model3_phase2}
\begin{eqnarray}
\Phi_1(\bm r)=
\tan^{-1}\left( \frac{y-\delta/2}{x} \right)-
\tan^{-1}\left( \frac{y-\delta/2-d_y}{x} \right),\nonumber\\\\
\Phi_2(\bm r)=
\tan^{-1}\left( \frac{y+\delta/2}{x} \right)-
\tan^{-1}\left( \frac{y+\delta/2-d_y}{x} \right),\nonumber\\
\end{eqnarray}
\end{subequations}
and densities
\begin{subequations} \label{den}
\begin{eqnarray}
\rho_1(\bm r) & = & \frac{1}{2} \mathcal{N}(\bm{r})
\nu(x,y-\delta/2) \nu(x,y-\delta/2-d_y), \nonumber\\ \\
\rho_2(\bm r) & = & \frac{1}{2} \mathcal{N}(\bm{r})
\nu(x,y+\delta/2) \nu(x,y+\delta/2-d_y), \nonumber\\
\end{eqnarray}
\end{subequations}
where $\mathcal{N}(\bm{r})$ is the normalization factor to ensure
$\rho_1(\bm{r}) + \rho_2(\bm{r}) = 1$ and
$\nu(x,y)=(x^2 + y^2) / (x^2 + y^2 + w^2)$.

We set $\delta=1/\kappa$, and from the numerical results, the radius of
the vortex $w$ is estimated to be $2w \simeq 1/\kappa$.
Figure~\ref{f:M3E} shows the total energy as a function of $d_y$;
this is obtained by substituting Eq.~(\ref{eq:model3}) into
Eq.~(\ref{eq:E}).
It can be seen in Fig.~\ref{f:M3E} that local energy minima appear on
either side of the global minimum and form the energy barriers, that
stabilize the vortex-antivortex pair.
We note that without SOC, there are no such barriers, as can be seen
in Fig.~\ref{f:M3E}(a) for $\kappa=0$.
We also note that the barriers do not appear for uniform densities $\rho_1
= \rho_2 = 1/2$, and the inhomogeneous densities of Eq.~(\ref{den}) are
necessary for the barriers to form.
Hence, we conclude that this is the combined effect of SOC and the
nonlinear interaction.

\begin{figure}[tb]
\includegraphics[width=8.5cm]{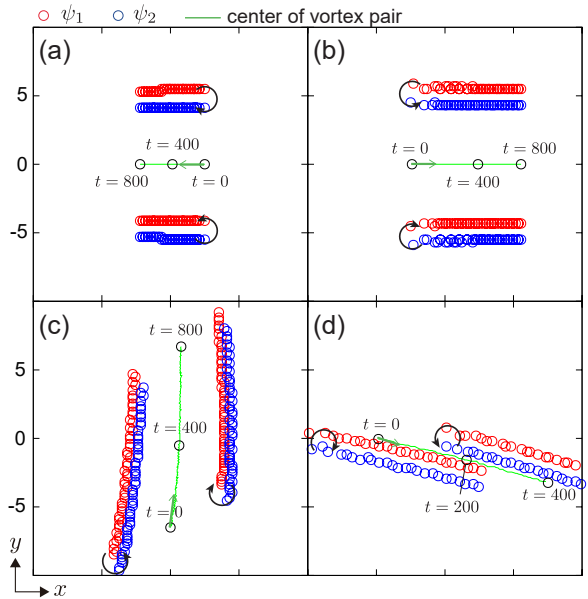}
\caption{
Trajectories of vortex-antivortex pairs for $\kappa=1$ and $\gamma = 0.8$.
Red and blue circles indicate the positions of the vortex cores in
$\psi_1$ and $\psi_2$, respectively.
The directions of the circulations of the vortices are indicated by black
arrows.
The initial distance between vortices is $d = 10$.
Black circles indicate the center of the vortex pairs
$(x_c,y_c)$ at $t=0$, $400$, and $800$, and green arrows indicate
the direction of motion.
See the Supplemental Material for movies of the dynamics~\cite{SM}.
}
\label{f:VAVP_dynamics}
\end{figure}

We now turn our attention to the dynamics of the vortex-antivortex pair.
Figure~\ref{f:VAVP_dynamics} shows the trajectories of the vortex cores,
where the initial state is prepared as follows.
We first prepare the state in Eq.~(\ref{eq:SV_INIT}) with the phase in
Eq.~(\ref{Phi}), and then we allow the imaginary-time evolution for a
short period (typically, $t \simeq 80$).
From this state, the real-time evolution begins.
Figures~\ref{f:VAVP_dynamics}(a) and \ref{f:VAVP_dynamics}(b) show the
dynamics of the vertically aligned vortex pair; the distance $d \simeq
10$ is larger than that of the stationary states shown in
Fig.~\ref{f:VAVP}.
The vortex-antivortex pair moves in the $-x$ and $+x$ directions at
constant velocity with a fixed distance between vortices.
These directions for the propagation agree with those for a scalar BEC.
However, the velocities $v_x \simeq -0.006$ in
Fig.~\ref{f:VAVP_dynamics}(a) and $v_x \simeq 0.011$ in
Fig.~\ref{f:VAVP_dynamics}(b) are much slower than $v_x = 1 / d \simeq
0.1$, which is that seen in a scalar BEC for the same $d$.
Figures~\ref{f:VAVP_dynamics}(c) and \ref{f:VAVP_dynamics}(d) show the
cases of oblique and horizontal alignments.
The propagation directions of these vortex pairs are different from those
in a scalar BEC.
This can be understood by inspecting the velocity field shown in
Fig.~\ref{f:VF}.
For example, on the negative $x$-axis in the left-hand panel of
Fig.~\ref{f:VF}(b), the velocity field is towards the lower right,
which indicates that a vortex located on the left-hand side of the
counterclockwise vortex will feel a mass current in this direction.
Similarly, a vortex located on the right-hand side of the clockwise vortex
will feel a mass current towards the lower right; this results in the
dynamics shown in Fig.~\ref{f:VAVP_dynamics}(d).

\begin{figure}[tb]
\includegraphics[width=7.0cm]{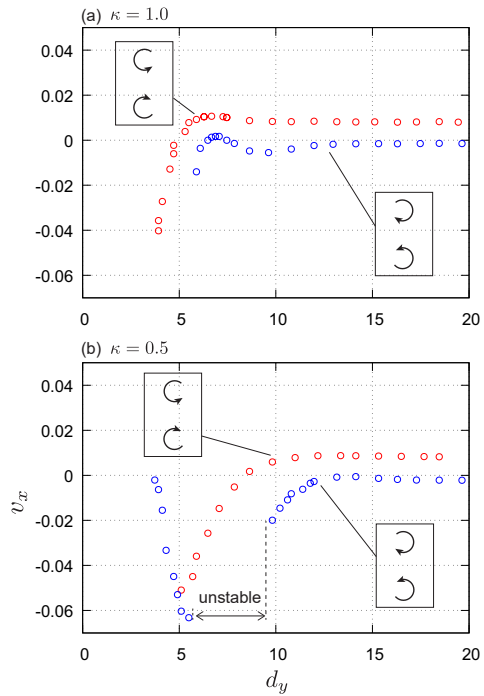}
\caption{
Velocity $v_x$ versus the distance $d_y$ of a vertically aligned
vortex-antivortex pair for (a) $\kappa=1$ and (b) $\kappa=0.5$ with
$\gamma = 0.8$.
The red and blue plots are for $\langle 1, -1 \rangle$ and $\langle -1,1
\rangle$, respectively.
The configurations of the vortex pairs are illustrated in the insets.
In (b), there is an unstable region (see text).}
\label{f:VAVP_v_d}
\end{figure}

Figure~\ref{f:VAVP_v_d} shows the velocity $v_x$ of the vertically aligned
vortex pair (i.e., $d_x = 0$) as a function of the vortex distance $d_y$,
which is obtained by a method similar to that used to obtain
Fig.~\ref{f:VAVP_dynamics}.
Such vortex pairs always travel in the $\pm x$ direction.
The $d_y$ dependence of the velocity is quite different from that in a
scalar BEC.
For $\kappa = 1$ (Fig.~\ref{f:VAVP_v_d}(a)), the velocity $v_x$ of the
$\langle 1, -1 \rangle$ pair (red circles) changes from negative to
positive as $d_y$ increases, and $v_x = 0$ at $d_y \simeq 5$, which
corresponds to the stationary state seen in Fig.~\ref{f:VAVP}(a).
The velocity $v_x$ of the $\langle -1, 1 \rangle$ pair (blue circles) also
crosses the $v_x = 0$ axis at $d_y \simeq 7$, which corresponds to the
stationary state seen in Fig.~\ref{f:VAVP}(b).
For $6 \lesssim d_y \lesssim 8$, the velocity changes from negative to
positive and from positive to negative as $d_y$ increases.
For a relatively large distance between vortices ($d \gtrsim 10$), the
propagation directions are the same as those of a scalar BEC, but the
$d_y$ dependence of $v_x$ is weak; this can be understood from the fact
that the velocity field is almost uniform far from the vortex core, as
shown in Fig.~\ref{f:VF}.
The velocity $|v_x|$ is always smaller than that in a scalar BEC for both
$\langle 1, -1 \rangle$ and $\langle -1,1 \rangle$ pairs.
There is no stable vortex-antivortex pair for small $d_y$;
the vortices are unstable against pair annihilation.
 
\begin{figure}[tb]
\includegraphics[width=8.0cm]{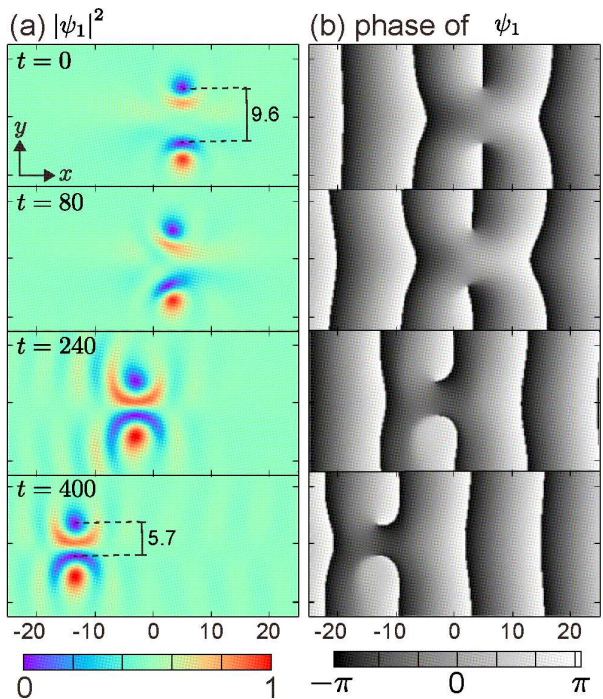}
\caption{Time evolution of the (a) density and (b) phase of the unstable
vortex-antivortex pair for $\kappa = 0.5$ and $\gamma = 0.8$, where the
vertical gauges indicate the distance $d_y$ between the vortex cores.
See the Supplemental Material for a movie of the dynamics~\cite{SM}.
}
\label{f:instability}
\end{figure}

The $\langle -1,1 \rangle$ pair exhibits interesting dynamics when
$\kappa$ is small.
As shown in Fig.~\ref{f:VAVP_v_d}(b), there is no stable $\langle -1,1
\rangle$ pair in the region $5.5 \lesssim d_y \lesssim 9.6$.
Figure~\ref{f:instability} shows the dynamics of the $\langle -1,1
\rangle$ pair with the initial distance $d_y = 9.6$, where the initial
state is prepared by the imaginary-time propagation for a short duration
from the initial phase in Eq.~(\ref{Phi}) with $d_y > 10$.
There is no stable state for $d_y = 9.6$ according to
Fig.~\ref{f:VAVP_v_d}(b).
As the vortex pair travels in the $-x$ direction, the distance $d_y$
decreases, and eventually the pair settles into a stable state with $d_y
\simeq 5.7$; the excess energy is released from the vortex pair as density
and spin waves.

\subsection{Vortex-vortex pair}
\label{s:VVP}

\begin{figure}[tb]
\includegraphics[width=8.0cm]{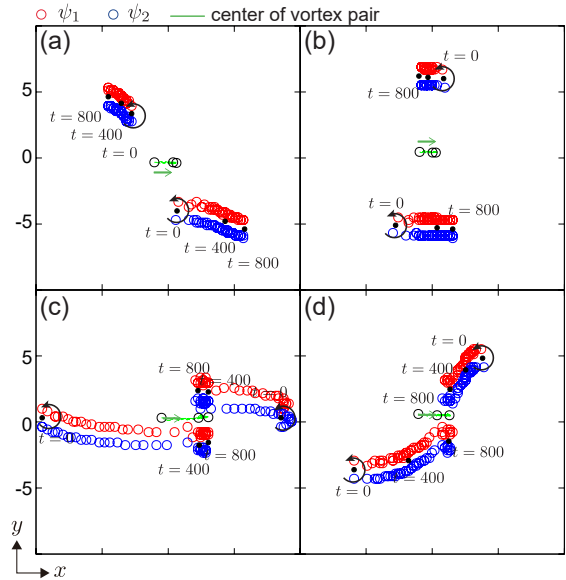}
\caption{Trajectories of vortex-vortex pairs for $\kappa=1$ and
$\gamma = 0.8$.
The initial vortex distance is (a) $d = 8.0$, (b) $d = 11.6$, (c) $d =
18.0$, and (d) $d = 12.9$.
Red and blue circles show the positions of the vortex cores in $\psi_1$
and $\psi_2$, respectively.
Black arrows indicate the direction of circulation.
Black circles are the center of the vortex pairs $(x_c,y_c)$ at
$t=0$, $400$, and $800$, and green arrows show the directions of motion.
See the Supplemental Material for a movie of the dynamics~\cite{SM}.
}
\label{f:VVP_dynamics}
\end{figure}

In a scalar BEC, two quantized vortices with the same circulation move
around each other.
In contrast, the dynamics of vortex-vortex pairs with SOC are
significantly different from those in a scalar BEC.
Figure~\ref{f:VVP_dynamics} shows the trajectories of vortices for the
$\langle 1, 1 \rangle$ pair, where the initial state is prepared by the
same method as in Fig.~\ref{f:VAVP_dynamics}.
When the initial positions are those shown in
Fig.~\ref{f:VVP_dynamics}(a), they move away from each other.
In the case of Fig.~\ref{f:VVP_dynamics}(b), the two vortices pass each
other.
The dynamics shown in Figs.~\ref{f:VVP_dynamics}(c) and
\ref{f:VVP_dynamics}(d) are more interesting.
The two vortices approach each other and unite to form a stationary state,
and the excess energy is released as waves.
The resultant stationary state is stable and remains at rest, and the
two vortices lie in a line perpendicular to the plane wave.
In all cases, the center of the pair initially moves in the direction of
$+x$.
The transformation in Eq.~(\ref{invert}) gives the dynamics of $\langle
-1, -1 \rangle$.

\begin{figure}[tb]
\includegraphics[width=8.0cm]{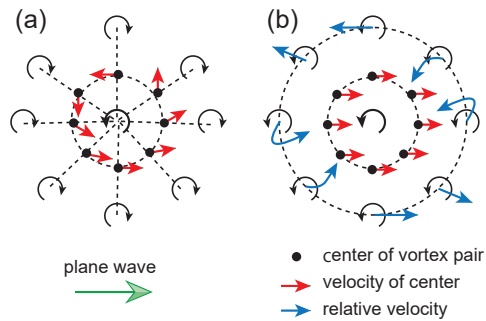}
\caption{
Schematic illustration of the dynamics of (a) vortex-antivortex pairs and
(b) vortex-vortex pairs when the distance is $d = 10$.
The red arrows indicate the direction of the velocity of $(x_c, y_c)$,
when one vortex is located at the origin.
In (a), the relative position of the vortices remains nearly constant.
In (b), the relative velocity (the velocity of the vortex in the moving
frame in which the other vortex is fixed to the origin) is indicated by
blue arrows.}
\label{f:direction}
\end{figure}

Figures~\ref{f:direction}(a) and \ref{f:direction}(b) summarize the
directions of the vortex motion when $d \simeq 10$ for the $\langle 1, -1
\rangle$ and $\langle 1, 1 \rangle$ pairs, respectively.
In Fig.~\ref{f:direction}(a), the motion of the center of the
vortex-antivortex pair $(x_c, y_c)$ is indicated by the red arrows, and
the relative position of the two vortices is nearly constant.
In Fig.~\ref{f:direction}(b), the relative motion of the vortex-vortex
pair is indicated by the blue arrow, and the center of the pair always
shifts in the direction of the plane wave.

\section{Conclusions}
\label{s:conclusions}
We have investigated the behaviors of quantized vortices in quasispin-1/2
BECs with Rashba SO coupling in a uniform 2D system, where the atomic
interactions satisfy the miscible condition and the ground state is the
plane-wave state.
We found that the static and dynamic properties of vortices are
significantly different from those of a scalar BEC.

For a single vortex state, we found that the vortex cores in two
components are shifted in the $\pm y$ directions by $\simeq 1 / \kappa$
(Fig.~\ref{f:SV}).
We also found that the phase distribution and velocity field around the
vortex are greatly deformed compared with those of a scalar BEC
(Figs.~\ref{f:SV} and \ref{f:VF}), which affects the dynamics of the
vortex pairs.
The vortex-antivortex pairs have stable stationary states at rest
(Fig.~\ref{f:VAVP}), and this is in marked contrast to the
vortex-antivortex pairs in a scalar BEC, which always travel.
The stationary states can be explained by variational analysis
(Fig.~\ref{f:M3E}).
Other than when in a stationary state, the vortex-antivortex pair travels
at a velocity much slower than that for a scalar BEC with the same vortex
distance.
The dependence of the velocity and moving direction on the vortex location
is also quite different from that in the case of a scalar BEC
(Figs.~\ref{f:VAVP_dynamics} and \ref{f:VAVP_v_d}).
The vortex-vortex pair exhibits interesting dynamics: the vortices
pass and move away from each other, or approach each other and combine
into a stationary state (Fig.~\ref{f:VVP_dynamics}).

In experiments, the vortex states shown in this paper may be produced by
the phase imprinting technique~\cite{S. Burger,J. Denschlag} and the
ensuing relaxation.
The dynamics of vortices can be observed by the destructive
imaging~\cite{Neely} or the real-time imaging~\cite{Freilich}.
We hope that our numerical results presented in this paper can provide
insight into a range of topics in the nonlinear dynamics of SO-coupled
BECs.

\begin{acknowledgments}
This work was supported by JSPS KAKENHI Grant Numbers JP16K05505,
JP26400414, and JP25103007, by the NMFSEID under Grant No. 61127901, and
by the Youth Innovation Promotion Association of CAS under Grant
No. 2015334.
\end{acknowledgments}

\end{document}